\documentclass[useAMS,usenatbib]{mn2e}

\usepackage{amsmath,amsfonts,amssymb}

\RequirePackage[dvips]{graphicx}

\usepackage{color}
\def\aap{\emph{A.\& A.}}

\def\apj{\emph{ApJ.}}
\def\ana{\emph{A\&A}}
\def\apjs{\emph{ApJ. Supp.}}
\def\apjl{\emph{ApJ. Lett.}}

\def\mnras{\emph{MNRAS}}

\title[The Hurst Exponent of Fermi GRBs]{The Hurst Exponent of Fermi GRBs}

\author[MacLachlan et al.]{G. A. MacLachlan$^{1}$\thanks{E-mail:
maclach@gwu.edu (GAM)}, A. Shenoy$^{1}$, E. Sonbas$^{2,3}$, 
R. Coyne$^{1}$, K. S. Dhuga$^{1}$, 
\newauthor A. Eskandarian$^{1}$, L. C. Maximon$^{1}$, and W. C. Parke$^{1}$\\
$^{1}$Department of Physics, The George Washington University, Washington, D.C. 20052, USA.\\
$^{2}$University of Adiyaman, Department of Physics, 02040, Adiyaman, Turkey.\\
$^{3}$NASA Goddard Space Flight Center, Greenbelt, MD 20771, USA.\\
}

\begin{document}

\maketitle

\label{firstpage}

\begin{abstract}
Using a wavelet decomposition technique, we have extracted the Hurst exponent 
for a sample of 46 long and 22 short Gamma-ray bursts (GRBs) detected by the
Gamma-ray Burst Monitor (GBM) aboard the Fermi satellite. 
This exponent is a scaling parameter that provides a measure of
long-range behavior in a time series. 
The mean Hurst
exponent for the short GRBs is significantly smaller than that for the
long GRBs. The separation may serve as an unbiased criterion for
distinguishing short and long GRBs.
\end{abstract}


\section{Introduction}\label{Introduction}

Our present understanding of complex astrophysical objects such as cataclysmic variables (CVs), active galactic nuclei (AGNs), 
and 
gamma-ray bursts (GRBs) comes nearly entirely from the temporal and spectral analyses of their photoemissions (with some additional information coming from
possible associations, such as host galaxies). 
In this paper we consider the temporal aspects of GRB light 
curves observed by the 
Gamma-Ray Burst Monitor (GBM) aboard the Fermi satellite. Many studies 
of the temporal properties of GRB light curves 
have been published, such as~\citet{Nemiroff00,Norris05,Hakkila09,Hakkila11,Nemiroff12}, 
from the perspective that light curves are comprised of a series of displaced pulses and that by fitting the individual pulses and associating pulses at various
photon energies one can arrive at a holistic understanding of light curves which in turn may be used to 
constrain the physics of the engines that produce them. The main appeal of this approach is the intuitive 
connection between pulses and collisions in the internal shock model. While this is a perfectly reasonable method, issues do 
arise concerning the functional form to use for pulse fitting and how to discern actual pulses from stochastic fluctuations 
in the light curves. The situation is further exacerbated by the fact that GRB light curves exhibit considerable 
variation in duration and in pulse profile. 
We note the significant progress made in non-parametric analyses using the Bayesian block technique~\citep{Scargle12}. In this type of analyis the duration of a light curve is represented as a tessellated block of time which can be partitioned into a complete array of sub-blocks in any number of ways. An optimal 
partition (a partition of sub-blocks maximizing a fitness function) is shown to exist, be unique, and computable iteratively. The optimal partition of sub-blocks is determined, given a prior probablility distribution for the number of blocks, by finding the model best representing the data as sets of piece-wise constant segments or sub-blocks. This technique shows great promise in resolving statistically significant
 temporal features from noise and detector related artifacts.

An ideal complementary approach to probing light curves would be one which handles seemingly disparate profiles on an equal 
footing and distills their complex forms into a single parameter which may be used to compare one light curve with another.  
One such method was pioneered by Harold Edwin Hurst~\citep{Hurst1951} with a technique he invented called the rescaled range analysis (R/S) 
which was later improved upon by Benoit Mandelbrot~\citep{Mandelbrot68}. 
The eponymous parameter resulting from the rescaled range analysis is 
called the Hurst exponent, $H$, and is closely related to the fractal dimension, D, the understanding of which Mandelbrot 
spent much of his career developing. In fact, fractional Brownian motion (fBm), which Mandelbrot defined in 1968, is parametrized 
solely by $H$ and serves as a useful model for discussing time series.  After determining $H$ for a given time series 
one is in a position to make several statements about the nature of that time series including whether the sequence appears 
random or whether it is persistent or anti-persistent, and if so, whether it exhibits long-range dependence, and over what 
time scales these characteristics are operative. All of these are informative quantitative statements, especially if the specific 
process generating the time series is partially or completely unknown, in which case, these statements are perhaps all one can really
say about the process given the available information. Some fields of research in which interesting work is being done with Hurst 
exponents are financial markets, seismology, anesthesiology, astrophysics, plasma physics and genomics.

We point out that neither the pulse fitting methods nor the 
Bayesian block analysis~\citep{Scargle12} yields 
information directly relatable to the Hurst exponent as does the 
wavelet analysis. One approach to access the Hurst exponent from 
a Bayesian block framework that seems reasonable would be an adaptation of
the Box-Counting algorithm~\citep{Feder1988}. Such a Bayesian-Box-Counting algorithm is outside
the scope of this paper.

The estimation of the Hurst exponent and the related scaling exponent, 
$\alpha$, has a history in 
astrophysics~\citep{Anzolin10,Tamburini09,Walker00,Fritz98}
for both Cataclysmic Variables (CVs) and GRBs.
We propose that a similar determination of $H$ for GRB light curves will be a valuable tool for categorization and we present 
a separation of long and short GRBs based on $H$.

\section[]{Methodology} \label{sec:method}

\subsection{Hurst Exponent and Self-Affinity}
\label{sec:selfaffn}
Pioneering work in self-similarity and long-range dependence was first 
published in 1951 by Hurst in the study of annual Nile River 
levels,~\citep{Hurst1951}. Hurst examined several 
decades of data to determine what should be the minimum size of a reservoir so that it neither overflows nor 
runs dry due to yearly fluctuations and made the unexpected observation that annual Nile River levels were 
not independent from one another but instead exhibited
a \emph{memory} of past events.

\begin{figure}
\includegraphics[width=84mm]{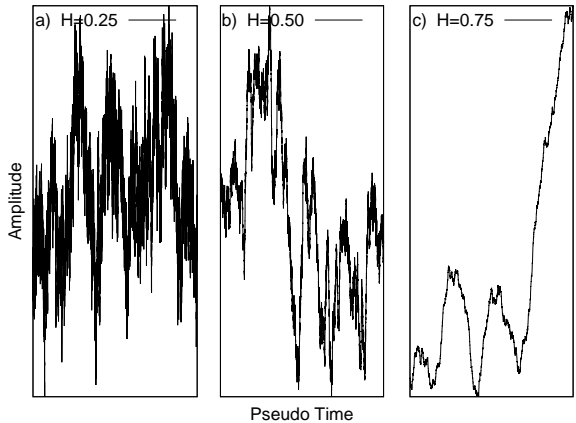}
\caption{ Simulated fractional Brownian motions with different values of $H$: a) $H=0.25$, b) $H=0.50$, c) $H=0.75$. }
\label{fig:LCS}
\end{figure}

\begin{figure}
\includegraphics[width=84mm]{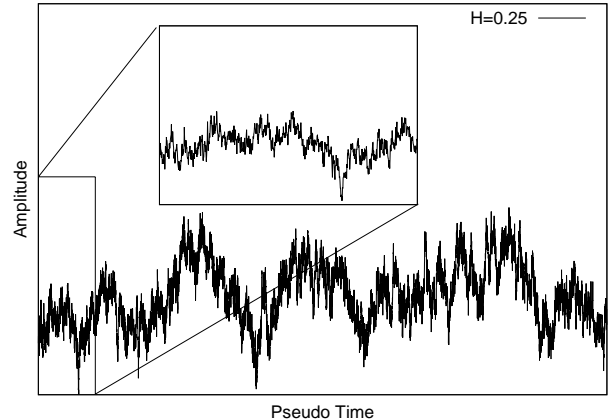}
\caption{Graph of $B_{H}(t)$ with $H=0.25$. A box is placed around a 
sub-range of $t$ (lower left hand corner). 
The box is zoomed into with time axis scaled 
by $a$ and amplitude scaled by  $a^{-H}$. This is a \emph{self-affine} 
transformation that not only makes the rescaled version qualitatively 
'similar' to the original but also preserves the variance as computed 
in Eq.~\ref{eq:fwt-variance}.   }
\label{fig:rescaleH0.25}
\end{figure}

\begin{figure}
\includegraphics[width=84mm]{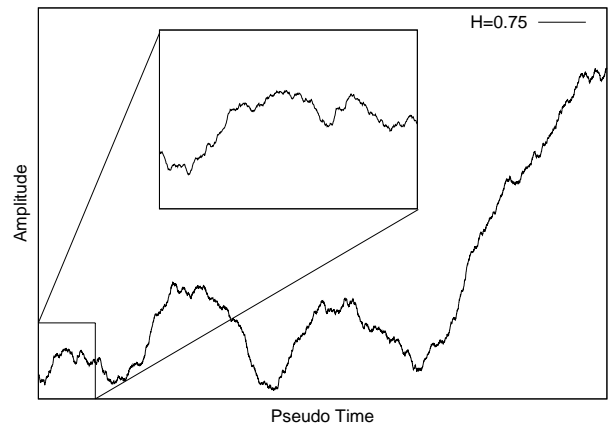}
\caption{Graph of $B_{H}(t)$ with $H=0.75$. A box is placed around a 
sub-range of $t$ (lower left hand corner). 
The box is zoomed into with time axis scaled 
by $a$ and amplitude scaled by  $a^{-H}$. This is a \emph{self-affine} 
transformation that not only makes the rescaled version qualitatively 
'similar' to the original but also preserves the variance as computed 
in Eq.~\ref{eq:fwt-variance}. }
\label{fig:rescaleH0.75}
\end{figure}

In this analysis of time-series data we search for statistical fractals, i.e., fractals whose statistical characteristics 
are independent of time scale. Such fractal time-series are called \emph{self-similar}. There is another class of 
statistical fractals whose scale invariance is broken but can be restored by a multiplicative 
factor. These statistical fractals are called \emph{self-affine}. 
\citet{Mandelbrot85} defined a time-series, $X(t)$ with $t\in\{t_0\mathellipsis t_{N-1}\}$, 
to be self-affine if, after a rescaling $t\rightarrow\lambda t$ the following 
relation is satisfied, 
\begin{equation}
X(t) \doteq \lambda^{-H}X(\lambda t).
\label{eq:self-affine-defined}
\end{equation}
The exponent, $H$, is the Hurst exponent,~\citep{Hurst1951}
and the symbol $\doteq$ denotes equality in 
distribution. The canonical example of a self-affine time-series, also given 
by \citet{Mandelbrot68}, is fractional Brownian motion, fBm.
Stationary in the context of this paper is second-order stationarity which 
means the first and second moments obey the following relations 
\begin{eqnarray}
\mathbb{E}\{X(t)\} & = & \mu_X   \nonumber \\
\mathbb{E}\{X(t_2)X(t_1)\} & = & \gamma(t_2-t_1)  =  \gamma(\tau),
\end{eqnarray}
where $\mu_X$ is the sample-mean, 
$\gamma$ is the auto-covariance sequence and $\tau\equiv t_2-t_1$ is 
the lag.
The Hurst exponent, $H$, parametrizes the degree of statistical self-similarity 
which a time-series exhibits. 
A self-similar series may be sub-divided into three categories: 
A series with $1/2<H<1$ is referred to as persistent or 
long-range dependent while 
a series with $0<H<1/2$ is referred to as anti-persistent,~\cite{Feder1988}. 
For $H=1/2$ we have neither persistence nor anti-persistence and this 
corresponds to the case of random and uncorrelated events. 
The Hurst exponent provides a model-independent characterization of the data.
Three examples of times series with different values of $H$ are shown  
in Fig.~\ref{fig:LCS}. A graphical depiction of the rescaling 
described by Eq.~\ref{eq:self-affine-defined} for a time series with $H=0.25$ 
is given in Fig.~\ref{fig:rescaleH0.25} and for a time series with 
$H=0.75$ in Fig.~\ref{fig:rescaleH0.75}.

\subsection{Wavelet Transforms}
Wavelet transformations have been shown to be a natural tool for multiresolution
analysis of non-stationary time-series \citep{Flandrin92,Mallat89}. 
Wavelet analysis is similar to 
Fourier analysis in many respects but differs in that a wavelet basis
function, $\psi(t)$, 
is well-localized while Fourier basis functions are global. 
Localization means that outside some range the amplitudes of wavelet basis 
functions go to zero or are otherwise negligibly small,~\cite{Percival00}. 
On the other hand,  the wavelet transform is similar to the Fourier transform because they both are expansions into a complete orthogonal basis and 
resolve low-frequency, large scale structure from high-frequency, 
small scale structure.

Wavelet analysis is said to be multiresolution because the time-series
under investigation is 
interrogated at multiple scales by a basis set of wavelets which are rescaled 
and translated versions of an original wavelet commonly referred to as 
the mother-wavelet, $\psi(t)$,
\begin{equation}
\psi(t)\rightarrow\psi_{a,b}(t)=\frac{1}{\sqrt{a}}\psi\left(\frac{t-b}{a}\right),
\label{eq:mother-wavelet-defined}
\end{equation}
where $a$ represents an octave or time-scale and the parameter $b$ 
gives the position of the wavelet within the octave. 

The continuous wavelet transform (CWT) 
coefficient, $C_{a,b}$, of a time-series for some
scale and position is computed as
\begin{equation}
C_{a,b}=\frac{1}{\sqrt{a}}\int X(t)\psi_{a,b}(t)dt.
\label{eq:wavelet-coeff}
\end{equation}

\paragraph{Wavelet Analysis}
The wavelet-transform technique for estimating self-affinity is outlined here.
By substituting the distribution relation in Eq.~\ref{eq:self-affine-defined} 
into Eq.~\ref{eq:wavelet-coeff} we find
\begin{eqnarray}
C_{a,b} & = & \frac{1}{\sqrt{a}}\int X(t)\psi_{a,b}(t)dt \nonumber
\\ & = & \frac{\lambda^{-(H+1/2)}}{\sqrt{\lambda a}}\int X(\lambda t)
\psi_{a,b}(\frac{\lambda t-\lambda b}{\lambda a})d(\lambda t) \nonumber
\\ & = & \lambda^{-(H+1/2)}C_{\lambda a,\lambda b}.
\label{eq:wavelet-self-affine}
\end{eqnarray}
It is straightforward to see from Eq.~\ref{eq:wavelet-self-affine}
that a self-affine time-series will have
wavelet coefficients whose variance 
over a particular scale, $\lambda a$, is related to the scale parameter 
$\lambda$ by,
\begin{equation}
\log {\rm var}(C_{\lambda a,\lambda b})=(2H+1)\log\lambda + {\rm constant}. 
\label{eq:cwt-variance}
\end{equation}

\paragraph{Fast Wavelet Transforms}
Similar to the CWT, the discrete fast wavelet transform (FWT) is also a multiresolution operation owing 
to the construction of the wavelets, $\psi_{j,k}$, which form the basis of the discrete fast  wavelet transform. 
We employed the discrete wavelet transform 
because of its high degree of computational efficiency. 
In order to distinguish between the CWT and its FWT counterpart we make a slight change of notation.
Just as before, the $\psi_{j,k}$, are rescaled, translated versions of the mother wavelet, $\psi$,
\begin{equation}
\psi_{j,k} = 2^{-j/2}\psi(2^{-j}t-k).
\end{equation}
The coefficients of the FWT are written as
\[
d_{j,k}=\langle X,\psi_{j,k}\rangle,
\]
where $j$ and $k$ play the roles of $a$ and $b$, respectively. Moreover, the
values which $j$ and $k$ assume obey the dyadic partitioning 
scheme~\citep{Mallat89,Addison02,Percival00}. That is,
for a time series whose number of elements is given by $N=2^m$,
\[
0\leq j\leq m-1,
\]
and
\[
0\leq k \leq 2^j-1.
\]
Applying the dyadic partitioning scheme removes any redundant encoding of 
information by the wavelet transform coefficients and guarantees 
orthogonality among the wavelet basis for any change in $j$ or $k$,
\begin{equation}
\langle \psi_{j,k},\psi_{j',k'}\rangle=\delta_{j,j'}\delta_{k,k'}.
\label{eq:dwt-ortho}
\end{equation}

\subsection{Logscale Diagrams}
The average power of the light curve at time scale $j$ is expressed as
$\beta_j$ and may be written in terms of the
variance of the FWT coefficients as
\begin{equation}
\beta_j={\rm var}(d_{j,k})=\frac{1}{n_j}\sum^{n_j-1}_{k=0}|d_{j,k}|^2,
\label{eq:fwt-variance}
\end{equation}
where $n_j$ is the number of coefficients at 
scale, $j$,~\citep{Abry00,Abry03}. Similarly to Eq.~\ref{eq:cwt-variance}, 
it has been shown by~\cite{Flandrin92} that 
for a series with non-stationary statistics  
the power-law variance of wavelet coefficients goes like
\begin{equation}
\log_2(\beta_j) = (2H+1)j+{\rm constant},
\label{eqn:hurst_est}
\end{equation}
where $H$ is the Hurst exponent.
\citet{Masry93} later extended this result to a larger class of 
non-stationary problems with stationary increments 
in the low-frequency limit and showed that fBms are a 
special case. A plot of Eq.~\ref{eqn:hurst_est}
is referred to as a logscale diagram.
Logscale diagrams are useful for identifying scaling 
regions, i.e., the range of octaves over which self-affine scaling occurs.
The slope, $\alpha$, of the scaling region is 
related to the Hurst exponent through $\alpha=2H+1$. 

In practice, a piecewise fitting function, $f(j;p_i)$ is defined,  
\[
 f(j;p_i) =
   \begin{cases}
     p_1 ;   1\leq  j\leq p_0\\ 
     p2+p_3j ;     p_0\leq j
   \end{cases}
\]
where $p_0$ is the value of $j$ at which the piecewise fitting function changes definition. 

\subsection{Choice of Wavelet Basis}
\label{sec:choice}
As in any orthogonal transformation, the basis functions to use in a wavelet transform is a matter of
strategic choice. One typically chooses a basis that emphasizes some characteristic of 
interest. Commonly used families of wavelet bases are the Coiflet,
Daubechies, and Haar~\citep{Addison02}. We chose the Haar wavelet basis which is the simplest of the 
Daubechies family.

The Haar wavelet basis was chosen from among all other possible bases because
it has the fewest number of vanishing moments and most compact 
support~\citep{Addison02}, 
has a straightforward interpretation, i.e., is equivalent to the Allan 
variance~\citep{Xizheng97} 
and is constant over its interval of support similar to the model
assumed in the Bayesian block method~\citep{Scargle98,Scargle12}.  

The Haar basis is not without some defects, as noted by~\citet{Kaplan93} 
and~\citet{Flandrin92}. Namely, the Haar wavelet transformation is known to underestimate
the actual Hurst exponent and this phenomenon is a function of the coarseness of 
the binning, the number of counts in the light curve, and also of H itself. We show in
Sec.~\ref{TestCase} that this effect is present but smaller than $\approx$1$\sigma$ for a set
of simulated light curves and is likely to be smaller for actual data. However, we consider that the 
advantages of the Haar basis outweigh its disadvantages.  

\subsection{Minimizing Uncertainties} 
\subsubsection{Circular Permutation}
Spurious artifacts
due to incidental symmetries resulting from accidental misalignment \citep{Percival00,Coifman95translation-invariantde-noising}
of light curves with wavelet basis functions are minimized by circularly
shifting the light curve against the basis functions. Circular shifting
is a form of translation invariant de-noising~\citep{Coifman95translation-invariantde-noising}.
It is possible a shift will introduce additional artifacts
by moving a different symmetry into a susceptible location. 
The best approach is to circulate the signal through all possible values,
or at least a representative sampling, and then take an average over the 
cases which minimizes the effect of   
spurious correlations.

\subsubsection{Reverse-Tail Concatenation}
\label{sec:rtc}
Both discrete Fourier and discrete wavelet transformations
imply that the expansion is periodic, with the longest period equal to 
the full time range of the input data.  This can be interpreted to mean 
that for a series of $N$ elements,
$\{X_0,X_1\mathellipsis X_{N-1}\}$ then $X_0$ is made a surrogate for $X_N$ and $X_1$ is made a surrogate for $X_{N+1}$, and so forth.
This assumption may lead to trouble if $X_0$ is much different from $X_{N-1}$.  In this case, artificially large
variances may be computed. 
Reverse-tail concatenation minimizes this problem by making a copy of 
the series
which is then reversed and concatenated onto the end
of the original series resulting in a new series with a length twice that 
of the original.
Instead of matching boundary conditions like,
\begin{equation}
X_0, X_1,\ldots,X_{N-1}, X_0,
\end{equation}
we match boundaries as,
\begin{equation}
X_0, X_1,\ldots X_{N-1},X_{N-1}, \ldots , X_1, X_0.
\end{equation}
Note that the series length has thus artificially been increased to $2N$ 
by reversing and doubling of the original series.
Consequently, the wavelet variances at the largest
scale in a logscale diagram reflect this redundancy. This is the reason that the 
wavelet variances at the
largest scale are excluded from least-squares fits of the scaling region.

\subsubsection{Poisson Operator}
\label{sec:poissonoper}
Photon counting statistics are considered in a bootstrapping procedure 
by applying a Poisson operator, $\mathcal{P}$($\lambda_i,X_i$), to 
every light curve prior to analyzing. Each light curve is binned 
initially at 200 
$\mu$-seconds and the number of counts per bin, $X_i$, 
is used as a mean value, 
$\lambda_i$, to be supplied to a 
Poisson random number generator. The value returned 
from $\mathcal{P}$($\lambda_i,X_i$) is used to replace the number of counts stored in $X_i$. The Poisson operator is applied to the signal $X_i$ 
prior to every circular permutation. We show in Sec.~\ref{TestCase} 
that
the Poisson operator does not affect the measured slope of 
logscale diagrams above the Poisson level.

\subsection[]{A Test Case: Fractional Brownian Motion}\label{TestCase}

Spatial-temporal fractional Brownian motions (fBm's) are a useful model 
for studying self-similarity and long-range dependence in 
non-stationary time-series,~\cite{Mandelbrot68} and are characterized by a 
single parameter, $H$, the Hurst exponent. 
An fBm with a particular $H$ is expressed as $B_H(t)$ and has the property of 
self-similarity over a range of scales after a rescaling of axes,
\begin{equation}
B_H(t)\doteq a^{-H}B_H(at),
\label{eqn:self-similar}
\end{equation}
where $\doteq$ denotes distributional equality as in Section~\ref{sec:selfaffn}.
The efficacy of the $H$ estimation procedure was tested using simulated data in the form of
fractional Brownian motion (fBm) time series.
Two tests were performed; in the first test we 
examine the ability of our algorithm to determine $H$ from fBms in the presence 
of Poisson noise 
and in the second test we examine how well we can determine $H$ at $H=0.25$, $H=0.50$,
and $H=0.75$ from noise-free fBms. 
\begin{figure*}
\includegraphics[width=168mm]{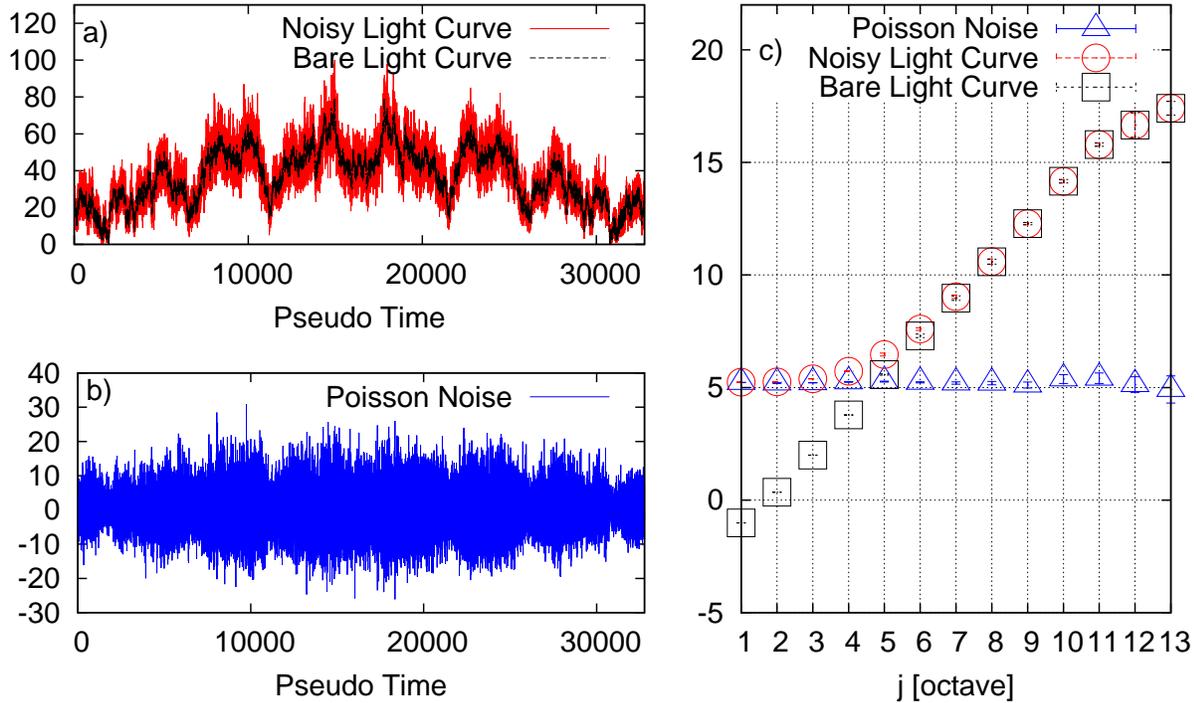}
\caption{Panel a) shows a sample fBm pre-processed and ready to be analyzed in black and the same light curve after applying the Poisson operator, $\mathcal{P}$, in red. Panel b) shows the Poisson noise that has been added by $\mathcal{P}$. In panel c) logscale diagrams illustrate the effect of Poisson statistics on the Hurst exponent. The bare fBm is shown in black, the dressed Poisson-type fBm is in red, and the residual Poisson noise in shown in blue. 
}
\label{fig:FBM_LC_LD}
\end{figure*}
\subsubsection{Test 1}
The numerical computing environment MATLAB
was used to produce 1000 realizations of fBms with scaling parameter $H$ randomly chosen from the range $0.0<\alpha<1.0$
by using a uniform random number generator. Copies of the 
fBms were combined with a
Poisson operator as described in Sec.~\ref{sec:poissonoper}.
The fBms and the Poissonian fBms thus produced are shown in black and red
respectively in panel a) of Fig.~\ref{fig:FBM_LC_LD}. Panel b) shows the Poisson noise that has been added by $\mathcal{P}$. The Logscale diagrams in panel c)
illustrate the effect of Poisson statistics on the Hurst exponent. 
The bare fBm is shown in black, the dressed Poisson-type fBm is in red, 
and the residual Poisson noise in shown in blue. 
The logscale diagram for the bare fBm in panel c) exhibits a clean slope 
across all octaves. We see the effect of a Poisson noise operator; it 
adds to the signal variance, constant across all octaves. Below some octave the signal
is completely dominated by noise but above that octave the slope of the logscale diagrams is independent of $\mathcal{P}$. See for example the black and red 
symbols for $j\geq6$.

\subsubsection{Test 2}

In the second test, 3000 simulated Poisson-type light curves were generated. The simulated data were divided into three subgroups of 1000 according to $H$. 
The three subgroups were $H=\{0.25,0.50,0.75\}$. The simulated data in each 
group were analysed and an attempt was made to recover the value of the Hurst
exponent, $H$, used to generate the fBm. The Hurst exponent was estimated by
a least squares fit to the scaling portion of the logscale diagrams to 
determine $\alpha$ and then $H$ is found from Eq.~\ref{eqn:hurst_est}.

\begin{figure}
\includegraphics[width=84mm]{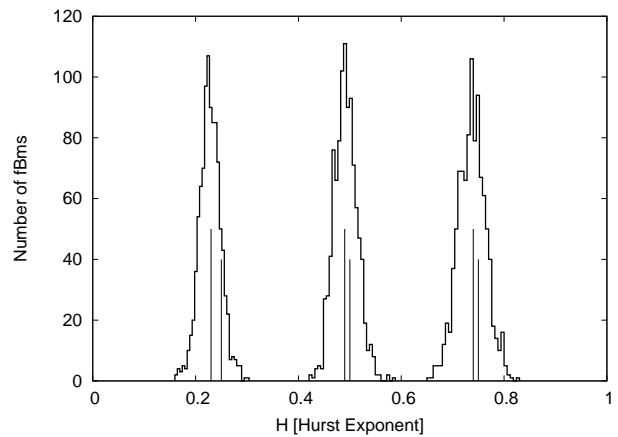}
\caption{Histograms of 3000 simulated fBm traces. Three categories of 
fBms were generated with known Hurst exponents, $H=0.25$, $H=0.50$, 
and $H=0.75$. These fBms were then analyzed to recover the Hurst 
exponent. The histograms are the results of this analysis. Pairs of vertical
lines are drawn for each peak. The shorter of the two indicates the known $H$
used to generate the fBms and the longer of the two indicates the $H$
extracted by our analysis. Results are tabulated in Table~\ref{tab:histo}. }
\label{fig:Hhisto}
\end{figure}

Results of the second test can be seen in Fig.~\ref{fig:Hhisto} and 
Table~\ref{tab:histo}.
The results show that the FWT analysis with the Haar wavelet basis does 
underestimate the value of $H$ as discussed in Sec.~\ref{sec:choice} 
but the magnitude of the error is not significant
for our purpose. 

\section{Data Reduction}\label{data}
The Gamma-Ray Burst Monitor (GBM) on board Fermi observes GRBs in the energy range 8\ keV to 40\ MeV. The GBM is composed of
12 thallium-activated sodium iodide (NaI) scintillation detectors (12.7 cm in diameter by 1.27 cm thick) that are
sensitive to energies in the range of 8 keV to 1 MeV, and two bismuth germanate (BGO) scintillation detectors (12.7 cm diameter
by 12.7 cm thick) with energy coverage between 200 keV and 40 MeV. The GBM detectors are arranged in such a way that they
provide a significant view of the sky \citep{Meegan09}.

In this work, we have extracted light curves for the GBM NaI detectors over the entire energy range (8 keV - 1 MeV, also
including the overflow beyond ~1 MeV). Typically, the brightest three NaI detectors were chosen for the extraction.
Lightcurves for both long and short GRBs were extracted at a time binning of 200 microseconds. The long GRBs were extracted
over a duration starting from 20 seconds before the trigger and up to about 50 seconds after the $T_{90}$
(taken from the Fermi GBM-Burst Catalog~\citep{Paciesas12})  for
the burst without any background subtraction. For short GRBs, durations were chosen to be
20 seconds before the trigger and 10 seconds after the $T_{90}$. The $T_{90}$ durations were obtained from the Fermi GBM-Burst Catalog~\citep{Paciesas12}. 
Summaries for the 46 long and 22 short GRBs used in this study are tabulated in Tables~\ref{tab:longs} and ~\ref{tab:shorts}.

\begin{table}
  \caption{Summary of results in Fig.~\ref{fig:Hhisto}.}
  \label{tab:histo}
  \begin{tabular}{@{}cc}
    \hline
     $H$ &
$H_{\rmn{meas}}$ \\
    \hline
     0.25 & 0.23$\pm$0.02 \\
     0.50 & 0.49$\pm$ 0.02\\ 
     0.75 & 0.74$\pm$ 0.03\\
    \hline
  \end{tabular}
  \medskip
\end{table}

\begin{table}
  \caption{Summary of results in Fig.~\ref{fig:HurstHisto}.}
  \label{tab:HurstHisto}
  \begin{tabular}{@{}lccc}
    \hline
     Type & N & STD & $\langle H\rangle$ \\
    \hline
     Long & 46 & 0.18 & 0.40$\pm$0.03 \\
     Short & 22 & 0.17 & 0.23$\pm$ 0.04\\ 
    \hline
  \end{tabular}
  \medskip
\end{table}

\begin{figure}
\includegraphics[width=84mm]{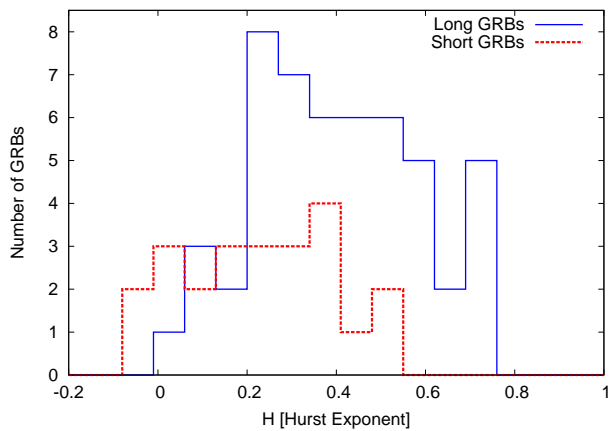}
\caption{Histogram of $H$ extracted from long and short GRBs. The result for 
long GRBs is plotted as the solid blue line while the short GRB result is plotted
with the dashed red line. Note the overlap but also that the means are displaced from one another
as shown in Tab.~\ref{tab:HurstHisto} for details of plot.}
\label{fig:HurstHisto}
\end{figure}

\section{Results and Discussion}
We have used a technique based on wavelets to determine the Hurst 
exponents for a sample of GRB prompt-emission light curves. 
As noted in Section~\ref{sec:selfaffn}, the Hurst exponent provides a measure 
of correlated 
behavior in a time series. The 
extreme values of $H$ vary from 0 to 1, and a value of 0.5 implies uncorrelated 
(random) behavior. As the fBm model indicates, large $H$ values tend to be 
associated with relatively smooth functions and small $H$ values tend to 
favor highly jagged curves. This feature suggests that $H$ may be useful in quantifying the
variability observed in GRB prompt-emission light curves. Plotted 
in Fig.~\ref{fig:HurstHisto} are the extracted $H$-exponents 
as histograms for both long and short GRBs. The histograms clearly show 
a displacement in $H$ for the distributions of long and short GRBs, 
with the short GRBs indicating a preference for small values of H (see Tab.~\ref{tab:HurstHisto}). 
The mean displacement in $H$ raises the interesting possibility of using 
this feature as a way of distinguishing between short and long GRBs. This would be in 
addition to the currently employed criteria based on $T_{90}$ and spectral 
hardness ratios. Interestingly, the histograms also show a significant overlap in the 
region of small $H$ exponents possibly signaling similarities between the two types of bursts in this range.

It could be argued that the sizable overlap of the distributions is essentially a consequence of the 
large dispersion (in $H$) exhibited by both short and long GRB distributions. While it is not known 
precisely what processes lead to this large dispersion in $H$, we note that the dispersion for the short 
GRBs is somewhat smaller than the corresponding one for long GRBs. If the dispersion is associated with the 
energetics of the progenitors of the respective systems, i.e., a merger of compact objects in the case of short 
GRBs and the collapse of a rapidly rotating massive star for long GRBs, then one might indeed expect 
a larger dispersion in the $H$-distribution of long GRBs compared to the corresponding one for short GRBs 
based purely on the difference in the mass range for the respective progenitors. Moreover, additional 
factors such as the formation of an accretion disk, the size of the disk, the mass of the disk, the 
strength of the magnetic field and the magnitude of the accretion rate during the prompt phase, remain 
largely uncertain. With the added intrinsic variability of the central engine itself, we should not be
surprised to observe a systematic difference in the extracted Hurst exponents for long and short bursts. 
For completeness, we mention that while the dispersion in $H$ is large for both distributions, the extracted $H$-value 
for each individual GRB is known reasonably precisely (see Table~\ref{tab:histo}). 

Another way to examine the $H$-distributions is to recast the data against the so-called minimum-time-scale parameter, MTS, 
extracted by \citet{MacLachlan12} and ~\citet{MacLachlan12a}. Using a method based on wavelets, these authors 
explored the scaling characteristics of GRBs and determined the minimum 
time scale at which scaling processes dominate over random noise processes. 
Furthermore, the authors have recently shown a direct connection between 
the extracted MTS and the smallest pulse structures extracted by pulse-fitting techniques. The same 
conclusions
were confirmed independently by \citet{Bhat13} using a similar technique to extract MTS by computing rescaled
Pearson variances. Furthermore, a link between pulse properties and MTS connecting GRB prompt 
emission and X-ray flaring has been identified by \citet{Sonbas13}. 

\begin{figure}
\includegraphics[width=84mm]{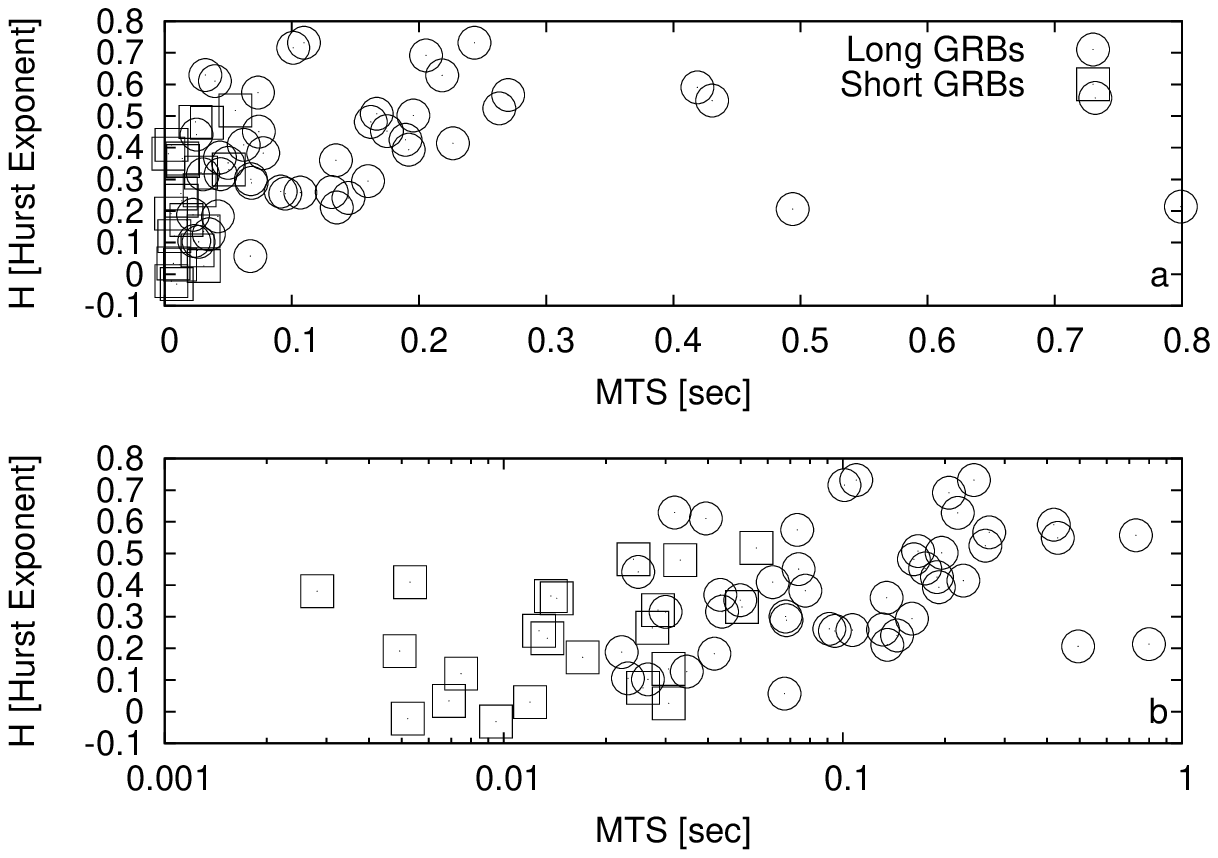}
\caption{A scatter plot of $H$ against the minimum variability time scale from~\citet{MacLachlan12} and~\citet{MacLachlan12a}. }
\label{fig:H_MTS}
\end{figure}

In addition to this link with pulses, MTS provides an alternate scale (to $T_{90}$) by which long and short GRBs can be separated. 
Shown in Fig.~\ref{fig:H_MTS} are the extracted $H$-exponents for both long and short GRBs versus the MTS (in the observer frame). 
Short GRBs tend to cluster around small MTS values and follow a steep trajectory in the $H$-MTS plane whereas 
the long GRBs are distributed over a larger range in MTS and seem to follow a gradual power-law-like trajectory.  
The behavior is a little more clear in panel (b) of Fig.~\ref{fig:H_MTS} where the MTS is plotted on a log 
scale: Here the the short and long GRBs indicate a small ($\sim30\%$) positive correlation respectively; the combined 
sample on the other hand shows a larger positive correlation ($\sim50\%$) and an obvious separation of the two distributions with MTS. 

Other astrophysical systems for which the Hurst exponent has been extracted includes CVs. These systems, 
comprising tightly-bound binaries (with periods of the order of few hours) and a primary consisting of a 
compact object (typically a white dwarf) and an accretion disk that can accommodate significant mass transfer 
from the secondary may provide a benchmark for gauging the systematics of the extracted Hurst exponents. Indeed, 
large dispersions in $H$ are found for both optical and X-ray light curves of CVs. Interestingly though, 
CVs apparently tend to favor large $H$-exponents i.e., greater than 0.5. This implies that the CV distributions are 
persistent as opposed to a tendency toward antipersistence for GRBs. By their very nature, CVs are systems that have built-in periodicity 
that is readily reflected, in most cases, in the observed emissions from these systems. GRBs, on the other hand, are transient phenomena 
which show very little evidence for periodicities. It's possible this simple difference may lead, in part at least, to the degree of 
persistence or antipersistence exhibited by these systems. Furthermore, some authors, \citep{Fritz98,Tamburini09,Anzolin10},
have noted that the extracted $H$'s indicate a sensitivity to the strength of the magnetic field of the systems under study, and in 
particular, the optical and x-ray emissions from CVs exhibit different $H$ distributions. Since the optical and x-ray emissions in CVs 
arise from spatially separated regions (the optical from an extended disk and the x-ray in the boundary layer between the inner 
regions of the disk and the surface of the compact object or the polar regions in the case of a highly magnetic system), it is tempting 
to surmise that such a comparison might be fruitful in elucidating the spatial characteristics of GRB jets: Examples
include the radii and or regions that are 
commonly associated with the emission sites for prompt gamma-rays (e.g., the photospheric radius in the case of a thermal component) 
and the steeply declining phase of the x-ray light curves (linked with high-latitude emission resulting from internal shell collisions). 
While it is understood that GRBs and CVs are very different systems and therefore the translation of the Hurst exponent from one 
system to the other is likely to be speculative at best, it is intriguing nonetheless that a simple scaling parameter may enable 
us to connect common underlying properties and processes that ultimately produce the observed emission 
in these diverse systems.

\section{Conclusions}
We have studied the temporal properties of a sample of prompt-emission 
light curves for short and long-duration GRBs detected by the 
Fermi/GBM mission. By using a technique based on wavelets we have 
extracted the Hurst exponents for these bursts. {\bf This exponent 
measures the relation between variability over the full range of available time-scales, comparing long-range with short-range variability.}
The physical limits of this 
index are 0 and 1, where the mid-point ($H = 0.5$), is an indicator of completely 
uncorrelated (random) processes that contribute to the observed time series. 
Often times, the $H$ is also associated with the fractal dimension ($D$) of structures by
\begin{equation}
D= 2-H,
\end{equation}
and can be thought of as a measure of the degree of jaggedness of the structures 
under study.  In this sense the $H$ may also be indirectly linked to the variability seen 
in the prompt-emission of many GRBs. Our main results 
are summarized as follows: 

a) The means of the $H$ distributions for the GRBs in our sample show an offset
between short and long GRBs, with the short GRBs indicating a preference for smaller Hurst exponents 
compared to the long GRBs. This offset is potentially an independent criterion for distinguishing between long and short-duration bursts.

b) Compared to short GRBs, long-duration bursts exhibit a larger dispersion in H. The origin of this 
dispersion is not known although it is possible that it is related to the underlying energetics of 
the different progenitors that produce long and short-duration bursts.

c) No distinct group or clustering is found for $H$ values corresponding to 0.5. This implies that random (or uncorrelated) 
processes, if present, play a lesser role in the production of the observed prompt emission. Moreover, the means of the 
$H$-distributions for both long and short GRBs indicate a skewness toward values less than 0.5. Overall, this implies that 
the prompt-emission time series exhibit antipersistence. 

Finally, we note that because of the large dispersion in $H$, there exists a significant region over which the long and short 
bursts overlap. This overlap region raises the interesting possibility of exploring bursts that may possess many more common 
features than would otherwise be suspected. The case for an intermediate class of GRBs~\citep{Horvath98,Gao10} remains unsettled and warrants 
further investigation.

\section{ACKNOWLEDGEMENTS}

The NASA grant NNX11AE36G provided partial support for this work and is gratefully acknowledged. 
The authors (GAM and KSD) acknowledge very useful discussions with Tilan Ukwatta.

\begin{table}
  \caption{Summary of Long GRBs.}
  \label{tab:longs}
  \begin{tabular}{@{}ccccc}
    \hline
     GRB Number & $H$ & $\delta H$ & $T_{90}$ [sec] &  $\delta T_{90}$ [sec] \\
    \hline
080723557 & 0.316 & 0.023 & 58.369 & 1.985 \\ 
080723985 & 0.425 & 0.053 & 42.817 & 0.659 \\ 
080724401 & 0.451 & 0.060 & 379.397 & 2.202 \\ 
080804972 & 0.549 & 0.085 & 24.704 & 1.460 \\ 
080806896 & 0.591 & 0.056 & 75.777 & 4.185 \\ 
080807993 & 0.105 & 0.014 & 19.072 & 0.181 \\ 
080810549 & 0.211 & 0.037 & 107.457 & 15.413 \\ 
080816503 & 0.258 & 0.035 & 64.769 & 1.810 \\ 
080817161 & 0.393 & 0.048 & 60.289 & 0.466 \\ 
080825593 & 0.382 & 0.036 & 20.992 & 0.231 \\ 
080906212 & 0.716 & 0.070 & 2.875 & 0.767 \\ 
080916009 & 0.414 & 0.053 & 62.977 & 0.810 \\ 
080925775 & 0.453 & 0.056 & 31.744 & 3.167 \\ 
081009140 & 0.732 & 0.073 & 41.345 & 0.264 \\ 
081101532 & 0.255 & 0.040 & 8.256 & 0.889 \\ 
081125496 & 0.629 & 0.080 & 9.280 & 0.607 \\ 
081129161 & 0.261 & 0.036 & 62.657 & 7.318 \\ 
081215784 & 0.629 & 0.070 & 5.568 & 0.143 \\ 
081221681 & 0.567 & 0.089 & 29.697 & 0.410 \\ 
081222204 & 0.502 & 0.065 & 18.880 & 2.318 \\ 
081224887 & 0.692 & 0.071 & 16.448 & 1.159 \\ 
090102122 & 0.126 & 0.013 & 26.624 & 0.810 \\ 
090131090 & 0.575 & 0.062 & 35.073 & 1.056 \\ 
090202347 & 0.241 & 0.039 & 12.608 & 0.345 \\ 
090323002 & 0.294 & 0.025 & 135.170 & 1.448 \\ 
090328401 & 0.289 & 0.034 & 61.697 & 1.810 \\ 
090411991 & 0.057 & 0.017 & 14.336 & 1.086 \\ 
090424592 & 0.442 & 0.029 & 14.144 & 0.264 \\ 
090425377 & 0.360 & 0.047 & 75.393 & 2.450 \\ 
090516137 & 0.206 & 0.026 & 118.018 & 4.028 \\ 
090516353 & 0.214 & 0.104 & 123.074 & 2.896 \\ 
090528516 & 0.259 & 0.026 & 79.041 & 1.088 \\ 
090618353 & 0.524 & 0.053 & 112.386 & 1.086 \\ 
090620400 & 0.508 & 0.052 & 13.568 & 0.724 \\ 
090626189 & 0.352 & 0.025 & 48.897 & 2.828 \\ 
090718762 & 0.482 & 0.055 & 23.744 & 0.802 \\ 
090809978 & 0.732 & 0.124 & 11.008 & 0.320 \\ 
090810659 & 0.558 & 0.104 & 123.458 & 1.747 \\ 
090829672 & 0.300 & 0.029 & 67.585 & 2.896 \\ 
090831317 & 0.102 & 0.013 & 39.424 & 0.572 \\ 
090902462 & 0.188 & 0.014 & 19.328 & 0.286 \\ 
090926181 & 0.369 & 0.032 & 13.760 & 0.286 \\ 
091003191 & 0.316 & 0.033 & 20.224 & 0.362 \\ 
091127976 & 0.611 & 0.060 & 8.701 & 0.571 \\ 
091208410 & 0.409 & 0.031 & 12.480 & 5.018 \\ 
100414097 & 0.183 & 0.020 & 26.497 & 2.073 \\ 

    \hline
  \end{tabular}
  \medskip
\end{table}

\begin{table}
  \caption{Summary of Short GRBs.}
  \label{tab:shorts}
  \begin{tabular}{@{}ccccc}
    \hline
     GRB Number & $H$ & $\delta H$ & $T_{90}$ [sec] &  $\delta T_{90}$ [sec]\\
    \hline
080723913 & 0.026 & 0.008 & 0.192 & 0.345 \\ 
081012045 & -0.022 & -0.002 & 1.216 & 1.748 \\ 
081102365 & 0.075 & 0.011 & 1.728 & 0.231 \\ 
081105614 & 0.135 & 0.026 & 1.280 & 1.368 \\ 
081107321 & 0.331 & 0.059 & 1.664 & 0.234 \\ 
081216531 & 0.366 & 0.046 & 0.768 & 0.429 \\ 
090108020 & 0.482 & 0.048 & 0.704 & 0.143 \\ 
090206620 & 0.358 & 0.042 & 0.320 & 0.143 \\ 
090227772 & 0.409 & 0.038 & 1.280 & 1.026 \\ 
090228204 & 0.381 & 0.027 & 0.448 & 0.143 \\ 
090308734 & 0.030 & 0.004 & 1.664 & 0.286 \\ 
090429753 & 0.321 & 0.045 & 0.640 & 0.466 \\ 
090510016 & 0.192 & 0.017 & 0.960 & 0.138 \\ 
090621922 & 0.255 & 0.053 & 0.384 & 1.032 \\ 
090907808 & 0.265 & 0.034 & 0.832 & 0.320 \\ 
091012783 & 0.120 & 0.014 & 0.704 & 2.499 \\ 
100117879 & 0.479 & 0.046 & 0.256 & 0.834 \\ 
100204858 & 0.518 & 0.070 & 1.920 & 2.375 \\ 
100328141 & 0.034 & 0.005 & 0.384 & 0.143 \\ 
100612545 & 0.171 & 0.021 & 0.576 & 0.181 \\ 
100625773 & 0.232 & 0.041 & 0.192 & 0.143 \\ 
100706693 & -0.031 & -0.009 & 0.128 & 0.143 \\ 

    \hline
  \end{tabular}
  \medskip
\end{table}


\begin{thebibliography}{99}

\bibitem[Abry et al.(2000)]{Abry00} Abry P., Flandrin P., Taqqu M. S., \& Veitch D., \ 2000, {\em Self-Similar Network Traffic and Performance Evaluation}, pg 39--88, New York: Wiley, K. Park and W. Willinger

\bibitem[Abry et al.(2003)]{Abry03} Abry P., Flandrin P., Taqqu M. S., \& Veitch D., \ 2003, {\em Theory and Applications of Long-Range Dependence}, Boston: Birkhauser, 527-556

\bibitem[Anzolin et al.(2010)]{Anzolin10} Anzolin G., Tamburini F., De Martino D., Bianchini A., \ 2002 , \ana, 519, A69

\bibitem[Addison(2002)]{Addison02} Addison P. S., \ 2002, \emph{The Illustrated Wavelet Transform Handbook}, IOP Publishing Ltd.

\bibitem[Bhat(2013)]{Bhat13} Bhat P.~N. \ 2013, arXiv:1301.4180v2 [astro-ph.HE].

\bibitem[Coifman(1995)]{Coifman95translation-invariantde-noising} Coifman R. R., Donoho D. L.\ 1995, Springer-Verlag, 125--150

\bibitem[Feder(1988)]{Feder1988} Feder J., \ 1988, \emph{ Fractals}, Plenum Press


\bibitem[Flandrin(1989)]{Flandrin89} Flandrin P.,\ 1989, IEEE, Transactions on Information Theory, 35, 197--199

\bibitem[Flandrin(1992)]{Flandrin92} Flandrin P.,\ 1992, IEEE, Transactions on Information Theory, 38, 910--917

\bibitem[Fritz \& Bruch(1998)]{Fritz98} Fritz T., Bruch A., \ 1998, \ana, 332, 586-604

\bibitem[Gao et al.(2010)]{Gao10} Gao H., Lu Y., \& Zhang S. N., \ 2010, \apj, 717, 268 

\bibitem[Hakkila \& Nemiroff(2009)]{Hakkila09} Hakkila J. \& Nemiroff R. J.\ 2009, \apj, 705, 372

\bibitem[Hakkila \& Preece(2011)]{Hakkila11} Hakkila J. \& Preese R.\ 2011, \apj, 740, 104

\bibitem[Horvath(1998)]{Horvath98} Horvath I., \ 1998, \apj, 508, 757

\bibitem[Hurst(1951)]{Hurst1951} Hurst H. ~E., \ 1951, Trans. Am. Sco. Civ. Eng., 116, 770

\bibitem[Kaplan \& Jay Kuo(1993)]{Kaplan93} Kaplan L. C., Jay Kuo C. C., \ 1989
, IEEE, Transactions on Signal Processing, 41, 3554--3562

\bibitem[MacLachlan et al.(2013)]{MacLachlan12} MacLachlan G. A., Shenoy A., Sonbas E., Dhuga K. S., Cobb B., Ukwatt
a T. N., Morris D. C., Eskandarian A., Maximon L. C., Parke W. C., \ 2013, \textit{Mon. Not. R. Astron. Soc}, doi: 10.1093/mnras/stt241

\bibitem[MacLachlan et al.(2012)]{MacLachlan12a} MacLachlan G. A., Shenoy A., Sonbas E., Dhuga K. S., Eskandarian A., Maximon L. C., Parke W. C.,\ 2012, \mnras~\emph{Letters}, 425, L32--L35

\bibitem[Mallat(1989)]{Mallat89} Mallat S. G.,\ 1989, IEEE, Transactions on Pattern Analysis and Machine Intelligence, 11, 674--693

\bibitem[Mandelbrot(1968)]{Mandelbrot68} Mandelbrot B. B. and Van~Ness J. W.\ 1968, SIAM Review, 10, 422-437

\bibitem[Mandelbrot(1985)]{Mandelbrot85}
        Mandelbrot B. B,
        SIAM Review,
        vol. 10,
        1968,
        pg 422--437

\bibitem[Masry(1993)]{Masry93}
        Masry E.,
        IEEE, Transactions on Information Theory,
        vol. 39,
        1993,
        pg. 260--264

\bibitem[Meegan et al.(2009)]{Meegan09} Meegan C. et al.\ 2009, \apj, 702, 791

\bibitem[Nemiroff(2000)]{Nemiroff00} Nemiroff R. J. et al.\ 2000, \apj, 544, 805

\bibitem[Nemiroff(2012)]{Nemiroff12} Nemiroff R. J.\ 2012, \mnras, 419, 1650 

\bibitem[Norris et al.(2005)]{Norris05} Norris J. P. et al.\ 2005, \apj, 627, 324

\bibitem[Paciesas et al.(2012)]{Paciesas12} Paciesas W.~S. et al.\ 2012, arXiv:1201.3099v1 [astro-ph.HE].

\bibitem[Percival(2000)]{Percival00} Percival D. B. and Walden A. T. \ 2002, \emph{Wavelet Methods for Time Series Analysis}, Cambridge University Press

\bibitem[Scargle(1998)]{Scargle98} Scargle J. D., \ 1998, 
	\apj, 
	504, 
	405--418

\bibitem[Scargle et al.(2012)]{Scargle12} Scargle J. D., Norris  J. P., Jackson B., Chiang J.,\ 2012, arXiv:1207.5578v2 [astro-ph.IM]

\bibitem[Sonbas et al.(2013)]{Sonbas13} Sonbas E., MacLachlan G. A., Shenoy A., Dhuga K. S., Parke W. C.,\ 2013, \apj, 767, L28

\bibitem[Tamburini et al.(2009)]{Tamburini09} Tamburini F., De Martino D., \& Bianchini A., \ 2009, \ana, 502, 1,1-5

\bibitem[Walker \& Schaefer(2000)]{Walker00} Walker K. C., Schaefer B.,\ 2000, \apj, 537, 264

\bibitem[Xizheng \& Zhensen(1997)]{Xizheng97} 
Xizheng K., Zhensen W.,\ 1993,
Frequency Control Symposium, 1997., Proceedings of the 1997 IEEE International,
515--518


\end{thebibliography}
\end{document}